\documentclass[aps,prl,amsmath,reprint,floatfix,superscriptaddress]{revtex4-1}
\usepackage{microtype} % improves pdf generation
\usepackage{amssymb}
\usepackage{graphicx}
\usepackage[usenames,dvipsnames]{xcolor} % text colors
\usepackage{enumerate}
\usepackage{bm}
\usepackage{bbold} % \mathbb{1}

\usepackage{times} % yields type setting that is more similar to the final APS version

\newcommand{\bvec}[1]{{\bm{#1}}} % bold vectors

\newcommand{\mi}{\mathrm{i}}
\newcommand{\mK}{{\mathrm{K}}}
\newcommand{\mL}{{\mathrm{L}}}
\newcommand{\mR}{{\mathrm{R}}}
\newcommand{\mH}{{\mathrm{H}}}
\newcommand{\Ac}{\mathcal{A}}
\newcommand{\loc}{{\mathrm{loc}}}
\newcommand{\coh}{{\mathrm{coh}}}
\newcommand{\imp}{{\mathrm{imp}}}
\newcommand{\hv}{{\mathrm{hv}}}
\newcommand{\lt}{{\mathrm{lt}}}
\newcommand{\io}{{\mathrm{io}}}
\renewcommand{\Re}{\mathrm{Re}\!\;}
\renewcommand{\Im}{\mathrm{Im}\!\;}

\RequirePackage[
  hyperindex,colorlinks,bookmarksnumbered,
  plainpages=true,pdfstartview=FitH]{hyperref}
\hypersetup{urlcolor={blue!50!black}, linkcolor={red!50!black},citecolor={blue!50!black}} 
\usepackage{hyperref}
\usepackage[all]{hypcap}

\begin{document} 

\title{Is the Orbital-Selective Mott Phase Stable against Interorbital Hopping?}
\author{Fabian B.~Kugler}
\affiliation{Department of Physics and Astronomy, Rutgers University, Piscataway, New Jersey 08854, USA} 
\author{Gabriel Kotliar}
\affiliation{Department of Physics and Astronomy, Rutgers University, Piscataway, New Jersey 08854, USA}
\affiliation{Condensed Matter Physics and Materials Science Department,\looseness=-1\,  
Brookhaven National Laboratory, Upton, New York 11973, USA}

\date{\today}

\begin{abstract}
The localization-delocalization transition is at the heart of strong correlation physics. Recently, there is great interest in multiorbital systems where this transition can be restricted to certain orbitals, leading to an orbital-selective Mott phase (OSMP). Theoretically, the OSMP is widely studied for kinetically decoupled orbitals, but the effect of interorbital hopping remains unclear. Here, we show how nonlocal interorbital hopping leads to local hybridization in single-site dynamical mean-field theory (DMFT). Under fairly general circumstances, this implies that, at zero temperature, the OSMP, involving the Mott-insulating state of one orbital, is unstable against interorbital hopping to a different, metallic orbital. We further show that the coherence scale below which all electrons are itinerant is very small and gets exponentially suppressed even if the interorbital hopping is not overly small. Within this framework, the OSMP with interorbital hopping may thus reach down to extremely low temperatures $T$, but not to $T=0$. Accordingly, it is part of a coherence-incoherence crossover and not a quantum critical point. We present analytical arguments supported by numerical results using the numerical renormalization group as a DMFT impurity solver. We also compare our findings with previous slave-spin studies.
\end{abstract}

\maketitle

The evolution of the electronic structure from localized to itinerant is a fundamental problem in condensed-matter physics and relevant to many interesting materials.
It continues to receive much experimental attention, as the transition region between localized and delocalized behavior hosts remarkable phenomena,
like high-temperature superconductivity \cite{Imada1998,Lee2006,Si2016}.

Recently, there has been a focus on multiorbital systems, triggered by the observation of orbital selectivity
whereby a subset of orbitals (denoted ``heavy'') has a much larger effective mass than another group (denoted ``light'').
An illustrative example under current study is FeTe$_{1-x}$Se$_x$
\cite{Yi2015,Liu2015,Otsuka2019,Huang2021}.
There, among the $t_{2g}$ orbitals, the $d_{xy}$ is the heaviest.
A central idea in this field is the orbital-selective Mott phase (OSMP) \cite{Anisimov2002},
where heavy electrons are Mott-localized and coexist with itinerant light electrons.
This idea is relevant to numerous model systems and materials \cite{Neupane2009,Vojta2010,Georges2013,Hardy2013,Pu2016,Miao2016,Lanata2017,Chen2019,Giannakis2019,Pascut2020,Kim2022}.
Often, a small difference among the orbitals at the one-particle level is drastically amplified by many-body correlations.
Importantly, a sharp localization-delocalization boundary can only be defined at zero temperature, 
$T \!=\! 0$,
via the participation of charge carriers in the volume of the Fermi surface.

The OSMP has been investigated intensively using dynamical mean-field theory (DMFT) \cite{Georges1996,Kotliar2006}
and slave-spin methods \cite{deMedici2005b,Hassan2010,Yu2012}.
There is consensus that the OSMP is realized within these methods in the absence of hopping matrix elements between different orbitals
\cite{Werner2009,deMedici2009,Kita2011,Huang2012,Wang2016,Kugler2019}.
This assumption is natural for local matrix elements 
(which are zero in high-symmetry situations 
\footnote{See Refs.~\cite{Koga2005,deMedici2005a,deMedici2005b,Komijani2017} for how local interorbital hybridization affects the OSMP.})
\nocite{Koga2005,deMedici2005a,deMedici2005b,Komijani2017}
but not for nonlocal ones
(which are allowed by symmetry) \cite{Slater1954}.
In realistic materials estimations,
the interorbital nonlocal hopping amplitudes are often comparable to those of the light electrons
\footnote{See, e.g., Table A.4 in Ref.~\cite{Haule2009} for iron pnictides and Tables S1 and S2 in the Supplemental Material of Ref.~\cite{Yu2018} for FeSe.}. 
\nocite{Haule2009,Yu2018}

\begin{figure}[b]
\includegraphics[scale=1]{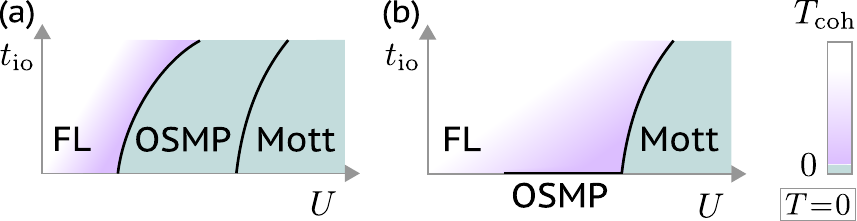}
\caption{%
Two possible scenarios, (a) and (b), for the zero-temperature ($T \!=\! 0$) phase diagram 
of multiorbital systems as a function of Coulomb repulsion $U$ and interorbital hopping $t_\io$.
Here, we provide evidence for scenario (b)
in which any finite $t_\io$ replaces the OSMP with a Fermi liquid (FL).
The coherence scale $T_\coh$, below which all electrons are itinerant,
is very low close to the OSMP and Mott phase.}
\label{fig:phase_diagram}
\end{figure}

Earlier attempts to study the OSMP in the presence of interorbital hopping $t_\io$ resulted in different pictures.
Using slave spins, Refs.~\cite{Yu2013,*Yu2017,Komijani2017} concluded that the OSMP survives finite $t_\io$ at $T \!=\! 0$,
while LDA+DMFT calculations of FeTe using a Monte Carlo impurity solver \cite{Yin2012}
\footnote{LDA+DMFT had previously proven successful for iron pnictides and chalcogenides \cite{Yin2011}, and nonlocal correlations beyond single-site DMFT were later shown to be weak \cite{Semon2017}.}
\nocite{Yin2011}\nocite{Semon2017}
argued for a smooth crossover,
where localization occurs only at sufficiently high $T$
\footnote{Refs.~\cite{Song2005,*Song2009,Ni2021} argue for a stable OSMP including interorbital hopping using DMFT and a Lanczos impurity solver. They consider a two-orbital model with an identical real-space dependence of all hoppings and use a canonical transformation that kinetically decouples the orbitals. This is, however, not the generic situation, and such a decoupling scheme is impossible once the intra- and interorbital hoppings have different momentum dependencies.}.
\nocite{Song2005,Song2009,Ni2021}
These two pictures are sketched in Fig.~\ref{fig:phase_diagram} as qualitative $T \!=\! 0$ phase diagrams.
They also lead to different behavior at finite temperature.
In the first case, one expects definite scaling behavior tied to a coherence scale $T_\coh$ which vanishes when a control parameter $x$ 
(e.g., interaction strength or doping) 
reaches a critical value $x_c$.
But the second scenario predicts a coherence-incoherence crossover where there is no such $x_c$ and $T_\coh$ stays finite.

Here, we settle this issue within the paramagnetic single-site DMFT in favor of the second scenario.
We provide analytic arguments why any finite $t_\io$ destabilizes the OSMP,
based on the DMFT equations.
The underlying mechanism has a 
simple physical interpretation,
and we show that the same mechanism is obstructed within the more approximate slave-spin methods
(thus explaining the results of Refs.~\cite{Yu2013,*Yu2017,Komijani2017}).
We obtain an exact numerical solution of the DMFT equations for a multiorbital model with interorbital hopping
using the numerical renormalization group (NRG) \cite{Bulla2008}.
This method is necessary to reach arbitrarily low $T$
and to show that $T_\coh$, while always finite, can be extremely small.

The basic argument is that
the DMFT views correlated systems as a collection of atoms, 
each of which hybridizes with the environment given by the rest of the lattice.
The low-energy hybridization plays a key role:
it is generically finite for Fermi liquids and vanishes for Mott insulators.
We will show that the low-energy hybridization of an electron in any orbital
is finite as long as it can hop to another, delocalized orbital and back.
This process is described by the momentum-dependent
interorbital hopping $\epsilon^\io_\bvec{k}$ and the momentum- and frequency-dependent
density of states $\Ac^\lt_{\bvec{k}\nu}$ of a light orbital,
as $\sum_\bvec{k} (\epsilon^\io_\bvec{k})^2 \Ac^\lt_{\bvec{k}\nu}$.
It is this low-energy hybridization which destabilizes the Mott state in favor of a Fermi-liquid ground state.
Below, we derive the hybridization formula for a two-orbital model,
discuss the coherence scale, and illustrate the consequences with numerical results.

\textit{Model.}---%
We consider a multiorbital Hubbard Hamiltonian
\begin{align}
\hat{H}
=
\textstyle
\sum_{ijnm\sigma} 
\hat{d}^\dag_{in\sigma} h^{nm}_{ij}  \hat{d}_{jm\sigma} 
+
\sum_i
\hat{H}_\mathrm{int}[\hat{d}_{in\sigma}]
,
\label{eq:Hamiltonian}
\end{align}
where 
$\hat{d}^\dag_{in\sigma}$
creates an electron at site $i$, in orbital $n$, and with spin $\sigma$.
The hopping matrix $h^{nm}_{ij}$ features nonlocal ($i \!\neq\! j$)
interorbital ($n \!\neq\! m$) hopping;
its Fourier transform is $h^{nm}_\bvec{k}$.
$\hat{H}_\mathrm{int}$ denotes the local interaction.
In single-site DMFT, correlations are assumed to be local \cite{Georges1996}.
The propagator reads 
$\bvec{G}_{\bvec{k}\nu} \!=\! [\nu \!+\! \mu \!-\! \bvec{h}_\bvec{k} \!-\! \bvec{\Sigma}_\nu]^{-1}$,
with the chemical potential $\mu$ and the retarded, matrix-valued self-energy $\bvec{\Sigma}_\nu$.
In sufficiently symmetric situations, 
one can choose the orbitals to be orthogonal, such that local one-particle objects are diagonal in orbital space
\cite{Kotliar2006,Poteryaev2008}.
This includes 
$\bvec{G}_{\loc,\nu} \!=\! \sum_\bvec{k} \bvec{G}_{\bvec{k}\nu}$,
$\bvec{\Sigma}_\nu$, and the on-site energies
$\bvec{\epsilon}_{d} \!=\! \sum_\bvec{k} \bvec{h}_\bvec{k} \!-\! \mu$.
Momentum sums are normalized: $\sum_\bvec{k} 1 \!=\! 1$.

A minimal model for the OSMP has two (orthogonal) orbitals,
a light ($\lt$) and a heavy ($\hv$) orbital. 
We write the general hopping matrix,
including the interorbital hopping $\epsilon^\io_\bvec{k}$, as
\begin{align}
\bvec{h}_\bvec{k} - \mu
=
\begin{pmatrix}
\epsilon^\lt_\bvec{k} & \epsilon^\io_\bvec{k} \\
\epsilon^\io_\bvec{k} & \epsilon^\hv_\bvec{k}
\end{pmatrix}
.
\label{eq:h-matrix}
\end{align}
The local propagator follows from a $2 \!\times\! 2$ matrix inversion as
\begin{align}
\bvec{G}_{\loc,\nu} 
& =
\sum_\bvec{k}
\frac{1}{ \prod_{n=\lt,\hv} [ \nu - \epsilon^n_\bvec{k}  - \Sigma^n_\nu ] - (\epsilon^\io_\bvec{k})^2  }
\nonumber
\\
& \ \times
\begin{pmatrix}
\nu - \epsilon^\hv_\bvec{k} - \Sigma^\hv_\nu & -\epsilon^\io_\bvec{k} \\
-\epsilon^\io_\bvec{k} & \nu - \epsilon^\lt_\bvec{k} - \Sigma^\lt_\nu
\end{pmatrix}
.
\label{eq:Gloc_matrix_inverse}
\end{align}
For our numerical results, we use the simplistic expressions 
\begin{align}
\epsilon^n_\bvec{k} 
& = - 2 t_n [ \cos(k_x) + \cos(k_y) + \cos(k_z) ] - \mu
, 
\nonumber
\\
\epsilon^\io_\bvec{k} 
& = -2 t_\io [ \cos(k_x) - \cos(k_y) ]
,
\label{eq:dispersion_relations}
\end{align}
for which diagonality of $\bvec{\epsilon}_d$ and $\bvec{G}_{\loc,\nu}$ 
is obvious.
However, our general arguments are independent of the choice of Eq.~\eqref{eq:dispersion_relations}.

\textit{DMFT equations.}---%
In DMFT, the lattice model is mapped onto an impurity model.
We call the (orbital-diagonal) impurity propagator
$\bvec{g}_{\nu} \!=\! [\nu \!-\! \bvec{\epsilon}_d \!-\! \bvec{\Delta}_\nu \!-\! \bvec{\Sigma}_\nu]^{-1}$,
where $\bvec{\Delta}_\nu$ is the retarded hybridization function.
The appropriate $\bvec{\Delta}_\nu$ is found by iteration until self-consistency
between the local lattice propagator and its impurity counterpart,
$\bvec{G}_{\loc,\nu} \!=\! \bvec{g}_{\nu}$,
is reached.

The diagonal elements of the local propagator are ($m \!\neq\! n$)
\begin{flalign}
G^n_{\loc,\nu} 
& =
\sum_\bvec{k}
\frac{1}{ r^n_{\bvec{k}\nu} - \Sigma^n_\nu }
,
\quad
r^n_{\bvec{k}\nu} 
= 
\nu - \epsilon^n_\bvec{k}
-
\frac{ (\epsilon^\io_\bvec{k})^2 }{ \nu - \epsilon^m_\bvec{k} - \Sigma^m_\nu }
,
\hspace{-0.5cm}
&
\label{eq:Gloc_explicit}
\end{flalign}
taken from Eq.~\eqref{eq:Gloc_matrix_inverse}.
The hybridization in the bare impurity propagator is then determined according to $G^n_{\loc,\nu} = g^n_{\nu}$.
With $1/g^n_{0,\nu} = \nu - \epsilon^n_d - \Delta^n_\nu$,
the value $\Delta^n_\nu$ can be found from 
\begin{align}
\frac{1}{g^n_{0,\nu}}
& = 
\Sigma^n_\nu + 
\frac{1}{G^n_{\loc,\nu}}
=
\frac{
\sum_\bvec{k} \frac{ r^n_{\bvec{k}\nu} }{ r^n_{\bvec{k}\nu} - \Sigma^n_\nu }
}{
\sum_\bvec{k} \frac{1}{ r^n_{\bvec{k}\nu} - \Sigma^n_\nu }
}
.
\label{eq:two-orbital_hybridization}
\end{align}
This intermediate result is key for the following discussion.
It gives the hybridization components for a general two-orbital system [Eq.~\eqref{eq:h-matrix}] according to the DMFT self-consistency condition. 
We reshuffled the self-energy from the numerator into the denominator,
but no approximation was made thus far.

While Eq.~\eqref{eq:two-orbital_hybridization} holds at self-consistency,
during the DMFT iteration, it is used to update
$\Delta^n_\nu$ from a given solution of the impurity model (yielding $\Sigma^n_\nu$) to the next.
We can briefly check the noninteracting case, $\Sigma^n_\nu = 0$, for which DMFT self-consistency is trivial.
There, Eq.~\eqref{eq:two-orbital_hybridization} correctly yields
$g^n_{0,\nu} = \sum_\bvec{k} \frac{1}{ r^n_{\bvec{k}\nu} }$.
Next, we use Eq.~\eqref{eq:two-orbital_hybridization} to investigate whether
the OSMP is stable against interorbital hopping.
To this end, we start from a converged DMFT solution with $t_\io = 0$, realizing the OSMP.
Then, we turn on $t_\io$ to check if the Mott insulator persists.

\begin{figure}[t]
\includegraphics[scale=1]{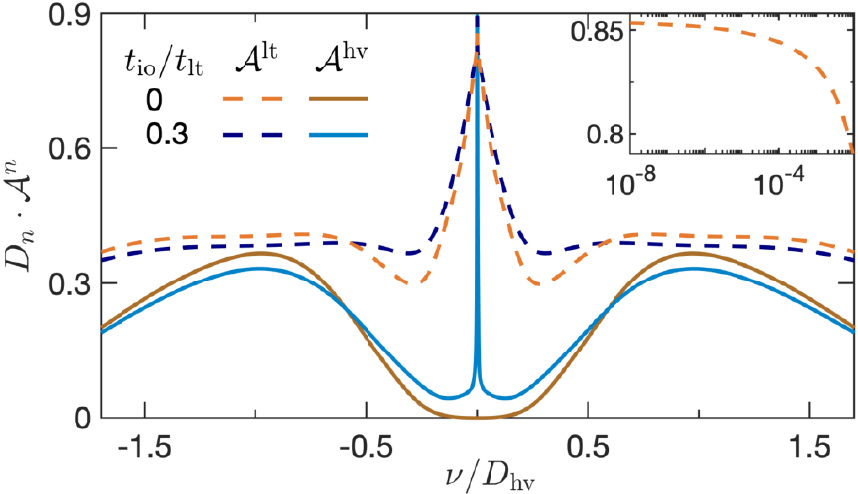}
\caption{%
Spectral functions $\Ac^n$ for the light and heavy orbitals.
At $t_\io \!=\! 0$, we find an OSMP with $\Ac^\hv$ gapped.
Finite $t_\io$ destabilizes the OSMP as
$\mathcal{A}^\hv$ develops a thin quasiparticle peak.
Inset: In the OSMP, $\lim_{\nu \to 0} \Ac^\lt_{\nu}$ converges only asymptotically \cite{Greger2013,Kugler2019}.}
\label{fig:Alin}
\end{figure}

\begin{figure*}[t]
\includegraphics[scale=1]{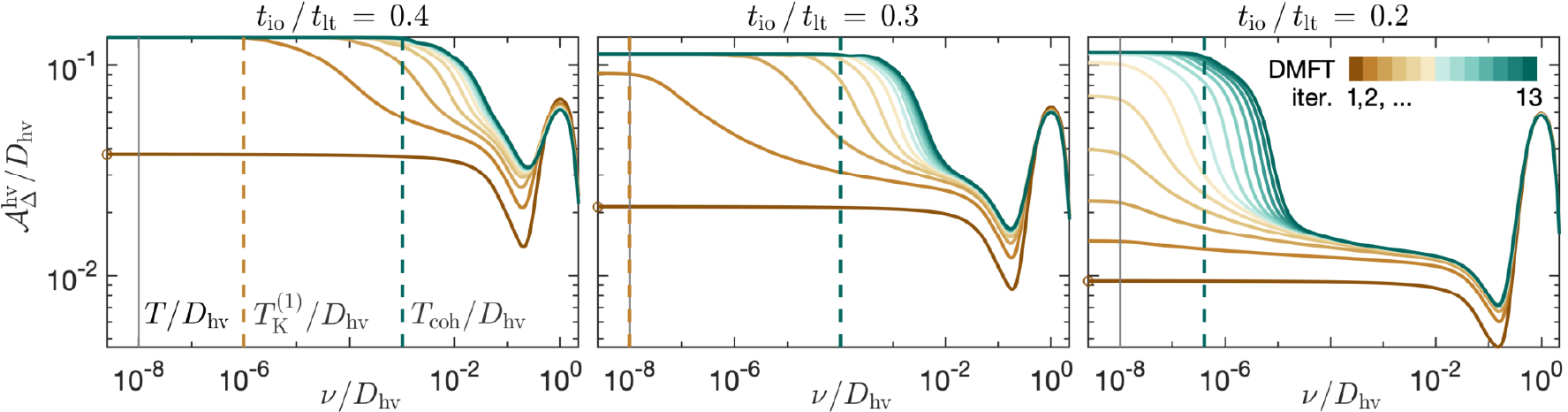}
\caption{%
Hybridization function $\Ac_\Delta$ in the heavy orbital for several DMFT iterations with finite $t_\io$,
starting from the OSMP solution at $t_\io \!=\! 0$.
Circles on the vertical axis give values $(t_\io/t_\lt)^2 \!\times\! \mathrm{const}$ according to Eq.~\eqref{eq:interorbital_hybridization}.
Dashed vertical lines indicate the coherence scale after the first DMFT iteration, $T_\mK^{(1)}$, 
and after the last DMFT iteration, $T_\coh$.
For $t_\io/t_\lt \!=\! 0.2$, $T_\mK^{(1)} \!\ll\! T$ and $T_\coh \!\gtrsim\! T$;
this opens a window of intermediate energies $|\nu| \!>\! T_\coh$ with OSMP-like features (cf.\ Fig.~\ref{fig:ASElog}).}
\label{fig:Hyblog}
\end{figure*}

Indeed, starting at $t_\io \!=\! 0$ and setting, e.g., $t_\hv \!\ll\! t_\lt$ at large interaction and half filling,
the heavy orbital is Mott insulating while the light orbital remains metallic.
The Mott insulator is signaled by a gap in the local density of states $\Ac^\hv_\nu$,
where $-\pi \Ac^n_\nu \!=\!\Im G^n_{\loc,\nu} \!=\! \Im g^n_\nu$,
and a divergent effective mass---i.e.,
$\lim_{\nu\to 0}|\Sigma^\hv_\nu| \!=\! \infty$.
The impurity solution yielding $\bvec{g}_\nu$ and $\bvec{\Sigma}_\nu$ is determined by
the hybridization $\bvec{\Delta}_\nu$ with spectral weights $\Ac^n_{\Delta,\nu} \!=\! - \Im \Delta^n_\nu / \pi$.
In most cases \cite{Hewson1993,deLeo2004,Georges2013,Aron2015,Walter2020},
a Fermi-liquid ground state is found if all $\Ac^n_{\Delta,\nu}$ are finite around $\nu \!=\! 0$,
while a Mott-insulating orbital requires a gapped $\Ac^n_{\Delta,\nu}$.

Now, we perform the first DMFT update, starting from the OSMP solution but setting $t_\io \!\neq\! 0$.
It is clear from Eq.~\eqref{eq:Gloc_explicit} that 
$\lim_{\nu\to 0}|\Sigma^\hv_\nu| \!=\! \infty$
makes $G^n_{\loc,\nu=0}$ for both $n$ independent of $\epsilon^\io_\bvec{k}$,
so that, in particular, $\Ac^\hv_\nu$ remains gapped. 
However, the result of the next iteration is determined
by $\Ac^n_\Delta$, not $\Ac^n$.
The divergent self-energy also simplifies the updated hybridization function.
In the limit $\nu \!\to\! 0$ within the OSMP, Eq.~\eqref{eq:two-orbital_hybridization} yields
\begin{flalign}
|\Sigma^\hv_\nu| \!\to\! \infty \! : \,\,\,
\frac{1}{g^\hv_{0,\nu}}
& = 
\sum_\bvec{k} r^\hv_{\bvec{k}\nu} 
,
\quad
\Delta^\hv_\nu
=
\sum_\bvec{k}
( \epsilon^\io_\bvec{k} )^2 
G^\lt_{\bvec{k}\nu}
.
\hspace{-0.5cm}
&
\label{eq:interorbital_hybridization}
\end{flalign}
For the second relation, $\nu - \epsilon_d^\hv$ in $1/g_{0,\nu}^\hv$
and $\sum_\bvec{k} (\nu - \epsilon_\bvec{k}^\hv)$ from $r_\bvec{k}^\hv$ cancel, and
$G_{\bvec{k}\nu}^\lt \!=\! 1/(\nu \!-\! \epsilon_\bvec{k}^\lt \!-\! \Sigma_\nu^\lt)$
when $\Sigma^\hv_\nu$ diverges.
Equation~\eqref{eq:interorbital_hybridization}
is our main result.
Assuming a divergent $\Sigma^\hv_\nu$
at low frequencies,
$\Delta^\hv_\nu$ retains a finite value,
independent of $\Sigma^\hv_\nu$.
Since a finite hybridization yields a Fermi-liquid ground state
in fairly general impurity models \cite{Hewson1993,deLeo2004,Georges2013,Aron2015,Walter2020},
we find that,
with $\Ac^\hv_{\Delta,\nu} = \sum_\bvec{k} (\epsilon^\io_\bvec{k})^2 \Ac^\lt_{\bvec{k}\nu} \!>\! 0$,
the Mott-insulating state of the heavy orbital
is unstable against interorbital hopping.
In further DMFT iterations, a quasiparticle peak in the heavy orbital will form, and $\Sigma^\hv_\nu$ will no longer diverge.
In the Supplemental Material \cite{Supp}, we show that Eq.~\eqref{eq:interorbital_hybridization} holds analogously for any number of orbitals, and we provide a free-energy functional to illustrate the universal nature of the effect described above.
\nocite{Weichselbaum2007,Kugler2020,Mitchell2014,Stadler2016,Kugler2022,Bulla1998,Kaufmann2019,Zitko2009,Kotliar1999,Kotliar2000,Bluemer2002,Loon2020}

We next include a temperature/energy coherence scale
below which the Fermi-liquid properties are found.
In the single-impurity Anderson model with large interaction $U$ and 
(nonsingular) hybridization $\Ac_\Delta$, 
this scale is the Kondo temperature $T_\mK \! \propto \! \exp( -\alpha U/\Ac_{\Delta,\nu=0})$
\footnote{Here, we use 
$T_\mK = 0.4107 \, (U\Gamma/2)^{1/2} \exp( - \tfrac{\pi U}{8\Gamma} + \tfrac{\pi\Gamma}{2U} )$,
where $\Gamma = \pi \Ac_{\Delta,\nu=0}$;
see Eq.~(6.109) ff.\ in Ref.~\cite{Hewson1993}.}.
\nocite{Hewson1993}
Similar behavior is expected for our model,
albeit with an effective $\tilde{U}$ encoding further microscopic parameters (like Hund's coupling $J$)
\footnote{Besides the local Hamiltonian, 
the shape of the DMFT hybridization affects the Kondo temperature \cite{Held2013}, too. 
Here, the only important aspect is that the exponential factor
$\exp [- 1/(\rho_0 J)]$ with $J \!\sim\! t^2/U$ remains the same.}.
\nocite{Held2013}
For the first DMFT iteration after switching from $t_\io \!=\! 0$ to $t_\io \!\neq\! 0$,
we have $\Ac^\hv_{\Delta,\nu=0} \!\propto\! t_\io^2 / t_\lt$
from Eq.~\eqref{eq:interorbital_hybridization}, and thus 
$T^{(1)}_\mK \!\propto\! \exp [ - \alpha \tilde{U} t_\lt / t_\io^2 ] = c^{(t_\lt/t_\io)^2}$,
with $c \propto \exp [ -\alpha \tilde{U}/t_\lt ]$ reminiscent of a single-orbital Kondo scale.
This shows that the coherence scale for the first DMFT iteration after the OSMP can be extremely small.
In the next iterations, 
$\Sigma^\hv_\nu$ no longer diverges,
and $\Ac^\hv_{\Delta,\nu=0}$ cannot be deduced as easily.
However, it is clear that delocalization of the heavy orbital will open more hybridization channels,
so that $T^{(1)}_\mK$ becomes a lower bound for 
the actual coherence scale after DMFT convergence, 
$T_\coh \!\geq\! T^{(1)}_\mK$.

\textit{Numerical results.}---%
We now turn to numerical results for the model of Eqs.~\eqref{eq:Hamiltonian}, \eqref{eq:h-matrix}, and \eqref{eq:dispersion_relations}.
We denote the half-bandwidth of $\epsilon^n_\bvec{k}$ by $D_n \!=\! 6 t_n$
and consider two half-filled orbitals with $D_\lt/D_\hv \!=\! 2$.
$D_\hv \!=\! 1$ is our energy unit, $T \!=\! 10^{-8}$, and
$\hat{H}_{\mathrm{int}}$ is given by the Kanamori Hamiltonian \cite{Supp}
with parameters $U \!=\! 2.4$ and $J \!=\! 0.4$ \cite{Georges2013}.
We use NRG as a real-frequency impurity solver for DMFT \cite{Supp} and assume paramagnetism.

To set the stage, Fig.~\ref{fig:Alin} shows two sets of spectral functions $\Ac^n$
for different interorbital hopping.
Our interaction parameters are such that $t_\io \!=\! 0$ realizes an OSMP,
where $\Ac^\hv$ has a gap, while $\Ac^\lt$ has a peak at $\nu \!=\! 0$.
Coupled to unscreened magnetic moments, the metallic orbital at $T \!=\! 0$
behaves as a singular Fermi liquid \cite{Greger2013,Kugler2019},
where $\lim_{\nu\to 0}\Ac^\lt_\nu$ converges only asymptotically (see inset)
and formally $Z_\lt \!=\! 0$.
For finite $t_\io$, $\Ac^\hv$ develops a narrow quasiparticle peak,
replacing the OSMP by a Fermi-liquid ground state.
Nevertheless, at larger energies $|\nu| \!\gtrsim\! 10^{-2}$,
the two sets of spectral functions for $t_\io \!=\! 0$ and $t_\io \!\neq\! 0$
are very similar.
Particularly, pronounced Hubbard bands in $\Ac^\hv$ exist in both phases
\footnote{Next to Hubbard bands, the insulating spectral function of the OSMP also features interband doublon-holon excitations (DHEs) \cite{Nunez2018,Kugler2019}. Using the techniques of Appendix B in Ref.~\cite{Kugler2019} here, the Hubbard bands occur at energies $\pm (U+J)/2 \!=\! \pm 1.4$ and the DHEs at $\pm 3J \!=\! \pm 1.2$, so that the DHEs are not discernible.}.
\nocite{Nunez2018}

Figure~\ref{fig:Hyblog} illustrates our instability argument.
It shows $\Ac_\Delta^\hv$ for several DMFT iterations with finite $t_\io$ starting from the OSMP solution at $t_\io \!=\! 0$.
In the first iteration, $\Ac^\hv_{\Delta,\nu} \!\propto\! (t_\io/t_\lt)^2$ according to Eq.~\eqref{eq:interorbital_hybridization}.
The resulting metallic state leads to an increased hybridization for the next iteration.
Its coherence scale (below which, e.g., the $\Ac^n_\nu$'s converge) roughly follows
$T^{(1)}_\mK \!\propto\! c^{(t_\lt/t_\io)^2}$.
For $t_\io/t_\lt \!=\! 0.2$, $T^{(1)}_\mK \!\ll\! T$, so that
$\Ac^\hv_{\Delta,\nu}$ for the next iteration converges below $T$ only.
In the subsequent DMFT iterations, the hybridization further builds up until the actual coherence
scale $T_\coh \!\geq\! T^{(1)}_\mK$ is established. 
For $t_\io/t_\lt \!=\! 0.2$, $T_\coh\!\gtrsim\! T$ is very low,
and OSMP-like behavior persists for $|\nu| > T_\coh$.

\begin{figure}[t]
\includegraphics[scale=1]{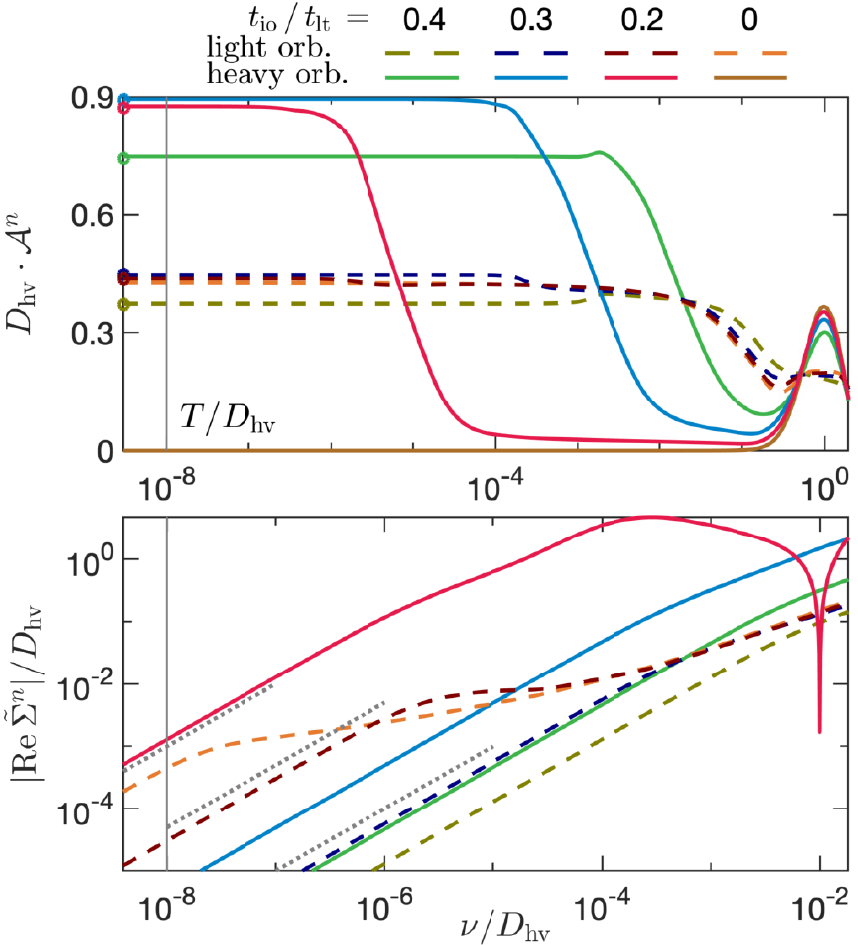}
\caption{%
Spectral functions $\Ac_\nu$ and self-energies
$\tilde{\Sigma}_\nu \!=\! \Sigma_\nu \!-\! \Sigma_{\nu=0}$
after DMFT convergence.
For $t_\io/t_\lt \!\in\! \{0.4, 0.3, 0.2\}$, Fermi-liquid behavior 
with $\Ac_{\nu=0}$ obeying Luttinger pinning (circles)
and linear $\Re\tilde{\Sigma}$ 
is seen below coherence scales 
of roughly $10^{-3}$, $10^{-4}$, and $4 \!\times\! 10^{-7}$, respectively
(cf.\ Fig.~\ref{fig:Hyblog}).
For $t_\io/t_\lt \!=\! 0.2$ and $10^{-5} \!<\! \nu/D_\hv \!<\! 10^{-1}$,
$\Ac^\hv$ almost vanishes, and
$\Re\tilde{\Sigma}^\lt$ perfectly follows
the logarithmic behavior
\cite{Greger2013,Kugler2019}
of the $t_\io \!=\! 0$ OSMP.
The three dotted lines indicate $\kappa\nu$,
with $\kappa \!=\! 10^{2}, 5 \!\times\! 10^{3}, 10^{5}$ from bottom to top.}
\label{fig:ASElog}
\end{figure}

Indeed, Fig.~\ref{fig:ASElog} shows the spectral functions and self-energies after DMFT convergence.
For all $t_\io \!>\! 0$, a Fermi-liquid ground state is obtained, with a finite quasiparticle peak 
obeying Luttinger pinning \cite{Mueller-Hartmann1989} $\Ac^n_{\nu=0} \!=\! \rho^n_{\nu=0}$
and self-energies having a linear real part.
At the lowest $t_\io \!>\! 0$, however, both properties are fulfilled only at very low energies
$|\nu| \!<\! T_\coh \!\sim\! 4 \!\times\! 10^{-7}$
(even though $\Ac^\lt$ increases most strongly around $|\nu| \!\sim	\! 10^{-1}$).
For $|\nu| \!>\! T_\coh$, the system is hardly distinguishable from the OSMP:
in an intermediate regime of around 4 orders of magnitude,
$\Ac^\hv_\nu$ almost vanishes and $\Sigma^\lt_\nu$ follows the logarithmic behavior
of the OSMP \cite{Greger2013,Kugler2019}.
While the quasiparticle weights $Z_n \!=\! 1/(1 - \partial_\nu \Re\Sigma^n_\nu|_{\nu=0})$
are already on the percent level for $t_\io/t_\lt \!=\! 0.3$, 
they reach values as low as $10^{-3}$ for the light and $10^{-5}$ for the heavy orbital 
at $t_\io/t_\lt \!=\! 0.2$.
Decreasing $t_\io$ further, an OSMP is recovered as $T_\coh \!<\! T \!=\! 10^{-8}$.

\textit{Comparison with slave spins.}---%
We finally compare our results to previous slave-spin studies, 
which found the OSMP to be stable against interorbital hopping \cite{Yu2013,*Yu2017}.
Slave-spin approaches decompose the physical fermions $\hat{d}$
into a bosonic slave-spin operator $\hat{b}$ and a slave fermion $\hat{f}^\dag$,
$\hat{d}^\dag_{im\sigma} \!=\! \hat{b}_{im\sigma} \hat{f}^\dag_{im\sigma}$.
It was shown that minimizing the mean-field decoupled free energy in slave-spin approaches
is equivalent to a DMFT-like treatment, where the slave-spin impurity solver yields the quasiparticle weight and DMFT self-consistency is imposed on the slave fermions (see Eq.~(30) in Ref.~\cite{Komijani2017}).

Expressions for the $f$ fermions follow from those of the $d$ fermions
by expanding the self-energy to linear order, $\Sigma^n_\nu \approx a_n + (1  - 1/Z_n) \nu$,
and dividing out $Z_n$.
For the local propagator [Eq.~\eqref{eq:Gloc_explicit}],
$G^{d,n}_{\loc,\nu} \!=\! Z_n G^{f,n}_{\loc,\nu}$,
this yields ($m \!\neq\! n$)
\begin{align}
G^{f,n}_{\loc,\nu} 
& =
\sum_\bvec{k}
\Big[
\nu - \epsilon^{f,n}_\bvec{k} 
-
\frac{ (\epsilon^{f,\io}_\bvec{k})^2 }{ \nu - \epsilon^{f,m}_\bvec{k} }
\Big]^{-1}
,
\\
\epsilon^{f,n}_\bvec{k}
& =
Z_n ( \epsilon^{d,n}_\bvec{k} + a_n )
,
\quad
(\epsilon^{f,\io}_\bvec{k})^2
=
Z_\lt Z_\hv (\epsilon^{d,\io}_\bvec{k})^2
.
\end{align}
Due to the factor $Z_\lt Z_\hv$,
the interorbital hopping has no effect here if $Z_\hv \!=\! 0$.
This agrees with our previous point
that a dominant $|\Sigma^\hv_\nu|$ makes 
$G^{d,n}_{\loc,\nu}$ independent of $\epsilon^{d,\io}_\bvec{k}$.
More insight is obtained from the impurity propagator,
$g^{d,n}_\nu \!=\! Z_n g^{f,n}_\nu$,
with
\begin{flalign}
g^{f,n}_\nu & \!=\! \frac{1}{\nu \!-\! \epsilon_f^n \!-\! \Delta^{f,n}_\nu}
, \quad
\epsilon_f^n \!=\! Z_n (\epsilon_d^n \!+\! a_n)
, \quad
\Delta^{f,n}_\nu \!=\! Z_n \Delta^{d,n}_\nu
.
\hspace{-0.5cm}
&
\end{flalign}
One finds that the $f$-fermion self-consistency condition
($G^{f,n}_{\loc,\nu} \!=\! g^{f,n}_\nu$)
leads to the same result for the $d$-fermion hybridization as in Eq.~\eqref{eq:interorbital_hybridization},
now in the form
$\Delta^{d,\hv}_\nu = \sum_\bvec{k} (\epsilon^{d,\io}_\bvec{k})^2 Z_\lt G^{f,\lt}_{\bvec{k}\nu}$.
This is still finite if the light orbital is metallic (here $Z_\lt \!>\! 0$).
However, the crucial difference is that the impurity model for the slave spins is not
characterized by $\bvec{\Delta}^d_\nu$ but by $\bvec{\Delta}^f_\nu$.
Here, each component is tied to the quasiparticle weight, $\Delta^{f,n}_\nu \!=\! Z_n \Delta^{d,n}_\nu$.
Hence, if $Z_\hv \!=\! 0$, the slave-spin impurity solver has no chance of seeing $\Delta^{d,\hv}_{\nu=0} \!\neq\! 0$
and, thereby, no chance of generating $Z_\hv \!>\! 0$ and leaving the OSMP.
In other words, the inseparable connection of $Z_n$ and $\Delta^{d,n}_\nu$ in slave-spin studies
leads to additional stationary points of the free energy, not present in DMFT.

\textit{Conclusion.}---%
Using single-site DMFT,
we showed that interorbital hopping $t_\io$ is a relevant perturbation to the OSMP
and destabilizes it at $T \!=\! 0$ in favor of a Fermi-liquid ground state.
The reason is that the low-energy hybridization in a given orbital 
has a finite contribution which stems from hopping to another orbital and back.
Crucially, this term depends only on the availability of states in the intermediate orbital
and not on the effective mass of the electron hopping.
While an arbitrarily large imbalance in effective masses can still exist,
within single-site DMFT, there is generically no OSMP with $t_\io \!>\! 0$ at $T \!=\! 0$,
and more generally below the coherence scale.
Its finite-temperature properties may thus be viewed as a coherence-incoherence crossover,
where selected orbitals are localized for $T \!>\! T_\coh$ but itinerant for $T \!<\! T_\coh$.
This crossover can either be tuned by increasing $T$ in a given system \cite{Yi2015}
or by decreasing $T_\coh$ at fixed (nonzero) $T$ (as in Ref.~\cite{Huang2021}).

Our analytic arguments are supported by numerical results using NRG
as a DMFT impurity solver, capable of accessing real frequencies and arbitrarily low temperatures.
This allowed us to demonstrate that $T_\coh$, below which
the Fermi-liquid properties are found, is very sensitive to system parameters 
and can be extremely small, even for moderate values of $t_\io/t_\lt$.
We showed that many properties of the $t_\io \!\neq\! 0$ state for energies above $T_\coh$
are almost indistinguishable from the $t_\io \!=\! 0$ OSMP that reaches down to $T \!=\! 0$.
Future theoretical work should aim to go beyond single-site DMFT
to address the influence of nonlocal, interorbital self-energy components
in renormalizing $t_\io$ \cite{DeLeo2008}.
Experimentally, our results can be tested by measuring the normal-state Fermi-surface volume at very low $T$
and by analyzing the scaling behavior in the OSMP at $T \!>\! 0$.

We thank A.~Gleis, K.~Haule, and Q.~Si for fruitful discussions 
and Seung-Sup B.~Lee and A.~Weichselbaum for a critical reading of the manuscript.
The NRG results were obtained using the QSpace tensor library developed by A.\ Weichselbaum 
\cite{Weichselbaum2012a,*Weichselbaum2012b,*Weichselbaum2020}
and the NRG toolbox by Seung-Sup B.\ Lee \cite{Lee2016,Lee2017}.
F.~B.~K.\ and G.~K.\ acknowledge support by NSF Grant No.\ DMR-1733071.
F.~B.~K.\ acknowledges support by the Alexander von Humboldt Foundation through the Feodor Lynen Fellowship. 

\textit{Note added.}---%
After completion of this work, we became
aware of Ref.~\cite{Stepanov2022}, which analyzes magnetic fluctuations
for the model we used to illustrate our findings.

\bibliographystyle{apsrev4-1}
\bibliography{references}

%%%%%%%%%%%%%%%%%%%%%%%%%%%%%%%%%%%%%%%%%%%%%%%%%

\clearpage

\setcounter{page}{1}
\thispagestyle{empty}

\onecolumngrid

\begin{center}
\vspace{0.1cm}
{\bfseries\large Supplemental Material for \\
``Is the Orbital-Selective Mott Phase Stable against Interorbital Hopping?''}\\
\vspace{0.4cm}
Fabian B.\ Kugler$^1$ and
Gabriel Kotliar$^{1,2}$
\\
\vspace{0.1cm}
{\it 
$^1$Department of Physics and Astronomy, Rutgers University, Piscataway, New Jersey 08854, USA\\
$^2$Condensed Matter Physics and Materials Science Department, Brookhaven National Laboratory, Upton, New York 11973, USA
} \\
\vspace{0.6cm}
\end{center}

\twocolumngrid

\begin{figure}[b!]
\includegraphics[scale=1]{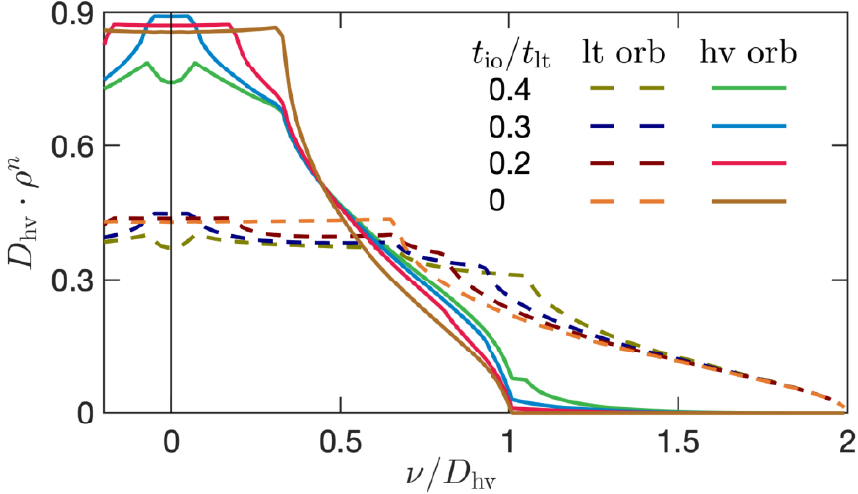}
\caption{%
The bare density of states $\rho^n$,
i.e.\ the noninteracting hybridization function $\Ac_\Delta^n$,
for various interorbital hoppings $t_\io/t_\lt$.}
\label{fig:DOS}
\end{figure}

In this Supplemental Material, we provide the computational details of our study, additional figures for the (particle-hole symmetric) bare density of states and the spectral functions away from particle-hole symmetry, and generalize the central Eqs.~\eqref{eq:two-orbital_hybridization} and \eqref{eq:interorbital_hybridization} of the main text to an arbitrary number of orbitals. Lastly, we discuss the multiorbital Mott transition by means of the DMFT free-energy functional. Citations refer to the list of references given in the main text.

\smallskip
\begin{center}
\small\bfseries COMPUTATIONAL DETAILS
\end{center}
\smallskip
We employ the full density-matrix (fdm) NRG \cite{Weichselbaum2007}
and exploit the
$\mathrm{U}(1)_\lt \times \mathrm{U}(1)_\hv \times \mathrm{SU(2)}_{\mathrm{sp}}$
charge and spin symmetry using the QSpace tensor library \cite{Weichselbaum2012a,*Weichselbaum2012b,*Weichselbaum2020}.
Here, we obtained an enlarged charge symmetry 
by neglecting the pair hopping of the Kanamori Hamiltonian
(without modifying the spin-flip part),
which can be justified \textit{a posteriori} if the probability of finding an empty and a fully occupied orbital is very low \cite{Kugler2020}.
For further efficiency, the Wilson chain of both orbitals are interleaved \cite{Mitchell2014,Stadler2016}.
We use an NRG discretization parameter of $\Lambda \!=\! 4$
and keep up to $25000$ SU(2)-spin multiplets in the beginning of the iterative diagonalization 
and around $15000$ at the end of it.
The self-energy is obtained with a novel equation-of-motion trick \cite{Kugler2022}
building on the established schemes of Refs.~\cite{Bulla1998} and \cite{Kaufmann2019}.
The resolution at finite energies is improved by
averaging results over two shifted discretization grids \cite{Zitko2009}
and by using an adaptive broadening scheme \cite{Lee2016,Lee2017}.
The DMFT iterations are performed until point-wise convergence of $\Ac_{\Delta,\nu}$ at the level of $10^{-4} D_\hv$ is reached.
Momentum sums are evaluated with more than $3 \!\times\! 10^7$ uniformly spaced momentum points in the irreducible Brillouin zone,
while manually setting $-\Im\Sigma^n_\nu \!\geq\! 0.004 D_\hv$ to obtain smooth curves.

\smallskip
\begin{center}
\small\bfseries ADDITIONAL FIGURES
\end{center}
%\smallskip
%
Figure~\ref{fig:DOS} shows the bare density of states for the system defined by
Eqs.~\eqref{eq:h-matrix} and \eqref{eq:dispersion_relations} of the main text
at $\mu=0$.
In a Fermi-liquid state,
Luttinger pinning \cite{Mueller-Hartmann1989} dictates
$\rho^n_{\nu=0} = \Ac^n_{\nu=0}$ at particle-hole symmetry.

Next, we discuss results away from particle-hole symmetry.
To this end, we refine Eq.~\eqref{eq:dispersion_relations} from the main text as 
\begin{align} 
\epsilon^n_\bvec{k} 
& = - 2 t_n [ \cos(k_x) + \cos(k_y) + \cos(k_z) ] - \mu_n
, 
\nonumber
\\
\epsilon^\io_\bvec{k} 
& = -2 t_\io [ \cos(k_x) - \cos(k_y) ]
,
\label{eq:dispersion_relationsNHF}
\end{align}
allowing for a crystal-field splitting $\Delta_{\mathrm{cfs}} = \mu_\lt - \mu_\hv$.
Indeed, while keeping equal hopping amplitudes $t_\lt = t_\hv$
(and thus bandwidths $D_\lt = D_\hv = 1$),
we use $\Delta_{\mathrm{cfs}} \neq 0$ to induce orbital differentiation 
and an imbalance in the effective masses.
For $\Delta_{\mathrm{cfs}} = 0.875$ and a suitable
average chemical potential $(\mu_\lt + \mu_\hv)/2$,
the heavy orbital is close to half filling whereas the light orbital is close to quarter filling.
Furthermore, we here choose interaction parameters $U=3$ and $J=0.5$.

\begin{figure}[b]
\includegraphics[scale=1]{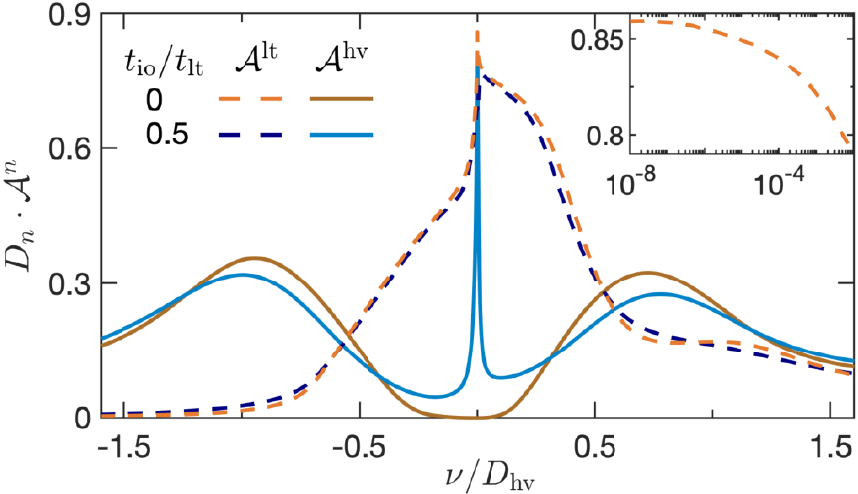}
\caption{%
Spectral functions $\Ac^n$ for the light and heavy orbital,
similarly as in Fig.~\ref{fig:Alin} of the main text, but without particle-hole symmetry.}
\label{fig:AlinNHF}
\end{figure}

Figure~\ref{fig:AlinNHF} presents our results away from particle-hole symmetry
in a way analogous to Fig.~\ref{fig:Alin} of the main text.
Without interorbital hopping, $t_\io \!=\! 0$, the system is an OSMP:
the spectral function of the heavy orbital is gapped while that of the light orbital
converges asymptotically to a finite value $A^\lt_{\nu=0}$ (see inset).
Again, the OSMP is unstable against interorbital hopping and,
for $t_\io/t_\lt \!=\! 0.5$, a quasiparticle peak in the heavy orbital develops.
Indeed, our analytic arguments on the instability of the OSMP against interorbital hopping 
given in the main text
are very general and independent of fillings or crystal fields.
We checked numerically that all statements regarding the build up of the hybridization in the heavy orbital, the low coherence scale $T_\coh$, and the OSMP-like properties
for $|\nu| \!>\! T_\coh$ also hold in this particle-hole asymmtric setup.

\newpage

\begin{center}
\small\bfseries DMFT SELF-CONSISTENCY FOR MANY ORBITALS
\end{center}
%\medskip

The central Eqs.~\eqref{eq:two-orbital_hybridization} and \eqref{eq:interorbital_hybridization} of the main text 
were derived by first rephrasing the two-orbital DMFT self-consistency condition
(as a quotient instead of a difference)
and then inserting a divergent self-energy component.
Here, we transform these results to the general matrix form for many orbitals.
First, we derive various expressions for the DMFT self-consistent hybridization function;
these might turn out helpful for future applications.
Second, we insert a divergent self-energy component in one of them to generalize
our DMFT characterization of the zero-temperature OSMP to an arbitrary number of orbitals.

Generally, the hybridization function can be viewed as a self-energy contribution.
Hence, there is a close analogy between the following formulas for $\bvec{\Delta}$
and the self-energy estimators derived in Ref.~\cite{Kugler2022}.
Still, as the hybridization arises from a quadratic coupling in the Hamiltonian,
the derivation is much simpler and amounts to a mere shifting of terms.
At first, we define a new local and thus orbital-diagonal variable 
\begin{align}
\bvec{S}_\nu
=
\nu - \bvec{\epsilon}_d - \bvec{\Sigma}_\nu
.
\label{eq:def_S}
\end{align}
On the impurity, it acts as an inverse ``bare propagator'' when 
$\bvec{\Delta}_\nu$ is viewed as the ``self-energy''. Indeed, we have
\begin{align}
\bvec{g}_\nu = ( \bvec{S}_\nu - \bvec{\Delta}_\nu )^{-1}
\end{align}
for the impurity propagator.
The local lattice propagator reads
\begin{align}
%\textstyle
\bvec{G}_{\loc,\nu} = \sum_\bvec{k} ( \bvec{S}_\nu - \tilde{\bvec{h}}_\bvec{k} )^{-1}
,
\end{align}
with the purely nonlocal hopping matrix
\begin{align}
\tilde{\bvec{h}}_\bvec{k}
=
\bvec{h}_\bvec{k} - \mu - \bvec{\epsilon}_d
, \qquad
%\textstyle
\sum_\bvec{k} \tilde{\bvec{h}}_\bvec{k}
= 0
.
\label{eq:htilde}
\end{align}
The DMFT self-consistency condition $\bvec{g}_\nu = \bvec{G}_{\loc,\nu}$ is then
\begin{align}
\bvec{\Delta}_\nu
& =
%\textstyle
\bvec{S}_\nu - \big[ \sum_\bvec{k} (\bvec{S} _\nu- \tilde{\bvec{h}}_\bvec{k})^{-1} \big]^{-1}
=
\bvec{S}_\nu - \bvec{G}_{\loc,\nu}^{-1}
.
\label{eq:DMFT-SC_matrix}
\end{align}

Now, Eq.~\eqref{eq:two-orbital_hybridization} in the main text,
solved for the hybridization, amounts to the rewriting
\begin{align}
\Delta^n_\nu
& =
S^n_\nu - \frac{1}{G_{\loc,\nu}^n}
=
%( S^n_\nu G_{\loc,\nu}^n - 1 ) \frac{1}{G_{\loc,\nu}^n}
%\\
%& =
%\frac{\sum_\bvec{k} S^n_\nu G_{\bvec{k}\nu}^{nn} - 1}{G_{\loc,\nu}^n}
%=
\frac{\sum_\bvec{k} (S^n_\nu - 1 / G_{\bvec{k}\nu}^{nn} ) G_{\bvec{k}\nu}^{nn} }{G_{\loc,\nu}^n}
.
\end{align}
For two orbitals, the matrix inversion in $\bvec{G}_{\bvec{k}\nu} = (\bvec{S} - \tilde{\bvec{h}}_\bvec{k})^{-1}$ 
can easily be done analytically
[$G_{\bvec{k}\nu}^{nn} = 1/(r^n_{\bvec{k}\nu} - \Sigma^n_{\nu})$,
cf.\ Eq.~\eqref{eq:Gloc_explicit} in the main text]
and the cancellation between $S^n_\nu$ and $1 / G_{\bvec{k}\nu}^{nn}$ used.
For an arbitrary number of orbitals, this is no longer the case, and we need to slightly adapt our approach.

To this end, we define the ``higher-order'' local correlators:
\begin{subequations}
\label{eq:higher-order_corr}
\begin{align}
\bvec{F}^\mR_{\loc,\nu}
& =
%\textstyle
\sum_\bvec{k} ( \bvec{S}_\nu - \tilde{\bvec{h}}_\bvec{k} )^{-1} \tilde{\bvec{h}}_\bvec{k}
,
\\
\bvec{F}^\mL_{\loc,\nu}
& =
%\textstyle
\sum_\bvec{k} \tilde{\bvec{h}}_\bvec{k} ( \bvec{S}_\nu - \tilde{\bvec{h}}_\bvec{k} )^{-1}
,
\\
\bvec{I}_{\loc,\nu}
& =
%\textstyle
\sum_\bvec{k} \tilde{\bvec{h}}_\bvec{k} ( \bvec{S}_\nu - \tilde{\bvec{h}}_\bvec{k} )^{-1} \tilde{\bvec{h}}_\bvec{k}
.
\end{align}
\end{subequations}
As local matrices, these are again orbital diagonal.
Consequently, 
$\bvec{F}^\mR_{\loc,\nu} = \bvec{F}^\mL_{\loc,\nu}$
follows directly from 
$\tilde{\bvec{h}}_\bvec{k} = \tilde{\bvec{h}}_\bvec{k}^\dag$.
Nevertheless, we keep the superscript for clarity.
Relations between the different correlators can be deduced with a simple trick,
by writing 
$\bvec{S}_\nu = \bvec{S}_\nu - \tilde{\bvec{h}}_\bvec{k} + \tilde{\bvec{h}}_\bvec{k}$,
and using
$\sum_\bvec{k} 1 = 1$ or
$\sum_\bvec{k} \tilde{\bvec{h}}_\bvec{k} = 0$.
For instance, we have
\begin{subequations}
\begin{align}
\bvec{G}_{\loc,\nu} \bvec{S}_\nu
& =
%\textstyle
\sum_\bvec{k} ( \bvec{S}_\nu - \tilde{\bvec{h}}_\bvec{k} )^{-1} \bvec{S}_\nu
\\
& =
%\textstyle
\sum_\bvec{k} ( \bvec{S}_\nu - \tilde{\bvec{h}}_\bvec{k} )^{-1} 
( \bvec{S}_\nu - \tilde{\bvec{h}}_\bvec{k} + \tilde{\bvec{h}}_\bvec{k} )
\\
& =
%\textstyle
1 + \sum_\bvec{k} ( \bvec{S}_\nu - \tilde{\bvec{h}}_\bvec{k} )^{-1} \tilde{\bvec{h}}_\bvec{k}
=
1 + \bvec{F}^\mR_{\loc,\nu}
\end{align}
\end{subequations}
and analogously 
$\bvec{S}_\nu \bvec{G}_{\loc,\nu} = 1 + \bvec{F}^\mL_{\loc,\nu}$.
Next, we also have
\begin{subequations}
\begin{align}
\bvec{S}_\nu \bvec{F}^\mR_{\loc,\nu}
& =
%\textstyle
\sum_\bvec{k} \bvec{S}_\nu ( \bvec{S}_\nu - \tilde{\bvec{h}}_\bvec{k} )^{-1} \tilde{\bvec{h}}_\bvec{k}
\\
& =
%\textstyle
\sum_\bvec{k}
( \bvec{S}_\nu - \tilde{\bvec{h}}_\bvec{k} + \tilde{\bvec{h}}_\bvec{k} )
( \bvec{S}_\nu - \tilde{\bvec{h}}_\bvec{k} )^{-1} 
\tilde{\bvec{h}}_\bvec{k}
\\
& =
%\textstyle
\sum_\bvec{k} \tilde{\bvec{h}}_\bvec{k} ( \bvec{S}_\nu - \tilde{\bvec{h}}_\bvec{k} )^{-1} \tilde{\bvec{h}}_\bvec{k}
=
\bvec{I}_{\loc,\nu}
\end{align}
\end{subequations}
and
$\bvec{F}^\mL_{\loc,\nu} \bvec{S}_\nu = \bvec{I}_{\loc,\nu}$.
With this, we can derive different estimators (distinguished by superscripts) for $\bvec{\Delta}$
from Eq.~\eqref{eq:DMFT-SC_matrix}:%
\begin{subequations}
\label{eq:Delta_estimators}
\begin{flalign}
\bvec{\Delta}^\mathrm{G}_\nu
& =
\bvec{S}_\nu - \bvec{G}_{\loc,\nu}^{-1}
,
&
\\
\bvec{\Delta}^\mathrm{G F}_\nu
& =
\bvec{G}_{\loc,\nu}^{-1} ( \bvec{G}_{\loc,\nu} \bvec{S}_\nu - 1 ) 
\,\, =
\bvec{G}_{\loc,\nu}^{-1} \bvec{F}^\mR_{\loc,\nu}
,
&
\\
\bvec{\Delta}^\mathrm{F G}_\nu
& =
( \bvec{S}_\nu \bvec{G}_{\loc,\nu} - 1 ) \bvec{G}_{\loc,\nu}^{-1}
\,\, =
\bvec{F}^\mL_{\loc,\nu} \bvec{G}_{\loc,\nu}^{-1}
,
&
\\
\bvec{\Delta}^\mathrm{I F}_\nu
& =
\bvec{F}^\mL_{\loc,\nu} \bvec{S}_\nu (\bvec{G}_{\loc,\nu} \bvec{S}_\nu)^{-1}
=
\bvec{I}_{\loc,\nu} (1 \!+\! \bvec{F}^\mR_{\loc,\nu})^{-1}
,
\hspace{-0.5cm}
&
\\
\bvec{\Delta}^\mathrm{F I}_\nu
& =
(\bvec{S}_\nu \bvec{G}_{\loc,\nu})^{-1} \bvec{S}_\nu  \bvec{F}^\mR_{\loc,\nu} 
=
(1 \!+\! \bvec{F}^\mL_{\loc,\nu})^{-1} \bvec{I}_{\loc,\nu} 
.
\hspace{-0.5cm}
&
\end{flalign}
Inspired by Ref.~\cite{Kugler2022}, we derive one more estimator as
\begin{align}
\bvec{\Delta}^\mathrm{I F G}_\nu
& =
\bvec{\Delta}^\mathrm{I F}_\nu ( 1 + \bvec{F}^\mR_{\loc,\nu} )
- \bvec{\Delta}^\mathrm{F G}_\nu \bvec{F}^\mR_{\loc,\nu}
\nonumber
\\
& =
( 1 + \bvec{F}^\mL_{\loc,\nu} ) \bvec{\Delta}^\mathrm{I F}_\nu 
- \bvec{F}^\mL_{\loc,\nu} \bvec{\Delta}^\mathrm{F G}_\nu 
\nonumber
\\
& =
\bvec{I}_{\loc,\nu}  - \bvec{F}^\mL_{\loc,\nu} \bvec{G}_{\loc,\nu}^{-1} \bvec{F}^\mR_{\loc,\nu}
.
\label{eq:Delta_IFG}
\end{align}
\end{subequations}
Interestingly, for $\bvec{\Delta}^\mathrm{I F G}_\nu$,
a $\bvec{k}$-independent shift of $\tilde{\bvec{h}}_\bvec{k}$
in each of the numerators of Eq.~\eqref{eq:higher-order_corr} cancels out.
If we define, e.g.,
$\tilde{\bvec{F}}^\mR_{\loc,\nu} = \sum_\bvec{k} \bvec{G}_{\bvec{k}\nu} \bvec{h}_\bvec{k}$,
$\tilde{\bvec{F}}^\mL_{\loc,\nu} = \sum_\bvec{k} \bvec{h}_\bvec{k} \bvec{G}_{\bvec{k}\nu}$, and
$\tilde{\bvec{I}}_{\loc,\nu} = \sum_\bvec{k} \bvec{h}_\bvec{k} \bvec{G}_{\bvec{k}\nu} \bvec{h}_\bvec{k}$,
then 
$\tilde{\bvec{\Delta}}^\mathrm{I F G}_\nu = 
\tilde{\bvec{I}}_{\loc,\nu}  - \tilde{\bvec{F}}^\mL_{\loc,\nu} \bvec{G}_{\loc,\nu}^{-1} \tilde{\bvec{F}}^\mR_{\loc,\nu} =
\bvec{\Delta}^\mathrm{I F G}_\nu$.
Finally, note that, since the local matrices are orbital diagonal, 
neither the distinction between 
$\bvec{F}^\mR_{\loc,\nu}$ and $\bvec{F}^\mL_{\loc,\nu}$
nor the order in the matrix products should play a role.

The higher-order estimators are convenient for extracting, e.g., the high-frequency behavior. 
From $\bvec{S}_\nu = \nu - \bvec{\epsilon}_d + \bvec{\Sigma}^\mH  + \mathit{O} \big( \nu^{-1} \big)$,
where $\bvec{\Sigma}^\mH$ is the constant Hartree part of the self-energy,
we get
\begin{subequations}
\begin{flalign}
\bvec{G}_{\loc,\nu}
& =
\nu^{-1} + \nu^{-2} \big( \bvec{\epsilon}_d \!+\! \bvec{\Sigma}^\mH \big)
+ \! \mathit{O} \big( \nu^{-3} \big)
,
&
\\
\bvec{F}_{\loc,\nu}
& =
\nu^{-2} \!
\sum_\bvec{k} \tilde{\bvec{h}}_\bvec{k}^2
+ \! \mathit{O} \big( \nu^{-3} \big)
,
&
\\
\bvec{I}_{\loc,\nu}
& =
\nu^{-1} \!
\sum_\bvec{k} \tilde{\bvec{h}}_\bvec{k}^2
+
\nu^{-2} \!
\sum_\bvec{k} \tilde{\bvec{h}}_\bvec{k} \big( \bvec{\epsilon}_d \!+\! \bvec{\Sigma}^\mH \big) \tilde{\bvec{h}}_\bvec{k}
+ \! \mathit{O} \big( \nu^{-3} \big)
.
\hspace{-0.5cm} &
\end{flalign}
\end{subequations}
Inserting these second-order expansions into Eqs.~\eqref{eq:Delta_estimators} yields
\begin{subequations}
\begin{flalign}
\bvec{\Delta}^\mathrm{G}_\nu
& =
\mathit{O} \big( \nu^{-1} \big)
,
&
\\
\bvec{\Delta}^\mathrm{F G}_\nu
& =
\nu^{-1} \!
\sum_\bvec{k} \tilde{\bvec{h}}_\bvec{k}^2
+ \! \mathit{O} \big( \nu^{-2} \big)
,
&
\\
\bvec{\Delta}^\mathrm{I F G}_\nu
& =
\nu^{-1} \!
\sum_\bvec{k} \tilde{\bvec{h}}_\bvec{k}^2
+
\nu^{-2} \!
\sum_\bvec{k} \tilde{\bvec{h}}_\bvec{k} \big( \bvec{\epsilon}_d \!+\! \bvec{\Sigma}^\mH \big) \tilde{\bvec{h}}_\bvec{k}
+ \! \mathit{O} \big( \nu^{-3} \big)
.
\hspace{-0.5cm} &
\label{eq:Delta_IFG_high-freq}
\end{flalign}
\end{subequations}
Hence, by virtue of $\bvec{\Delta}^\mathrm{IFG}$,
a simple expansion of the self-energy to the first, constant order
directly yields the first two high-frequency moments of the hybridization function.

The estimator $\bvec{\Delta}^\mathrm{IFG}$ in Eq.~\eqref{eq:Delta_IFG}
can also be phrased as
\begin{align}
\bvec{\Delta}_\nu
& =
\bvec{I}_{\loc,\nu}  - \bvec{\Delta}_\nu \bvec{G}_{\loc,\nu} \bvec{\Delta}_\nu
.
\label{eq:Delta_I_DGD}
\end{align}
This (implicit) form is useful to generalize Eq.~\eqref{eq:interorbital_hybridization} of the main text
to an arbitrary number of orbitals.
Let us again assume that the self-energy of one orbital $n$ (the ``heavy'' orbital) diverges at low energies,
$\lim_{\nu \to 0} |\Sigma^n_\nu| = \infty$.
Then, we can use Eq.~\eqref{eq:Delta_I_DGD} to 
expand the DMFT self-consistency condition for $\Delta^n_\nu$ as
\begin{subequations}
\begin{align}
\Delta^n_\nu 
& =
I^n_{\loc,\nu}  - \Delta^n_\nu G^n_{\loc,\nu} \Delta^n_\nu
\\
& =
\bigg[
\sum_\bvec{k} \tilde{\bvec{h}}_\bvec{k} \bvec{G}_{\bvec{k}\nu} \tilde{\bvec{h}}_\bvec{k} 
\bigg]_{nn}
+ 
\mathit{O} \big( 1 / S^n_\nu \big)
\label{eq:Delta_G_matrix}
\\
& =
\sum_\bvec{k}
\sum_{m, m' \neq n}
h_\bvec{k}^{nm} G_{\bvec{k}\nu}^{mm'} h_\bvec{k}^{m'n}
+ 
\mathit{O} \big( 1 / S^n_\nu \big)
.
\end{align}
\end{subequations}
For the last step, we used that 
$G_{\bvec{k}\nu}^{nn} = \mathit{O}\big( 1/S^n_\nu \big)$ and that
$\tilde{h}_\bvec{k}^{nm} \!=\! h_\bvec{k}^{nm}$ for $n \!\neq\! m$
since $\mu$ and $\bvec{\epsilon}_d$ in Eq.~\eqref{eq:htilde} are diagonal.
This is the desired generalization of Eq.~\eqref{eq:interorbital_hybridization} of the main text.

\medskip
\begin{center}
\small\bfseries FREE-ENERGY FUNCTIONAL
\end{center}
\medskip

To illustrate the universal nature of our DMFT description of the phase transition 
between Mott insulator and OSMP or Fermi liquid, we here provide the corresponding 
Landau free-energy functional \cite{Kotliar1999,Kotliar2000,Bluemer2002,Loon2020}.
We start from the general DMFT functional $\Gamma$ as given in Eq.~(83) of Ref.~\cite{Kotliar2006}. There, $\Gamma$ is a functional of the three variables 
$\bvec{G}_\loc$, $\bvec{\mathcal{M}}_{\mathrm{int}}$, $\bvec{\mathcal{G}}_0$. 
Here, we write $\bvec{S}$ [cf.\ Eq.~\eqref{eq:def_S}]
instead of $\bvec{\mathcal{M}}_{\mathrm{int}}$, and we use the hybridization function $\bvec{\Delta}$ instead of the bare propagator $\bvec{\mathcal{G}}_0$ (called $\bvec{g}_0$ here) as a variable. 
Thus, we have
\begin{flalign}
& \Gamma[\bvec{G}_\loc, \bvec{S}, \bvec{\Delta}] 
= 
F_\imp[\bvec{\Delta}]
- \mathrm{Tr}_{\mi\nu} [ \ln \bvec{G}_\loc ] 
&
\nonumber
\\
& \quad
- \mathrm{Tr}_{\mi\nu,\bvec{k}} \ln [ \bvec{S} - \tilde{\bvec{h}}_\bvec{k} ]
+ \mathrm{Tr}_{\mi\nu} [ (\bvec{S} - \bvec{\Delta} - \bvec{G}_\loc^{-1}) \bvec{G}_\loc ]
.
\hspace{-0.5cm} &
\label{eq:Gamma_DSG}
\end{flalign}
Note that $\bvec{G}_\loc$, $\bvec{S}$, $\bvec{\Delta}$ are matrices in orbital and spin space, depending on fermionic Matsubara frequencies $\mi\nu$. Accordingly,
\begin{align}
\mathrm{Tr}_{\mi\nu} [\,\cdot\,]
=
T \sum_{\mi\nu} \mathrm{Tr} [\,\cdot\,]
,
\quad
\mathrm{Tr}_{\mi\nu,\bvec{k}} [\,\cdot\,]
=
T \sum_{\mi\nu} \sum_\bvec{k} \mathrm{Tr} [\,\cdot\,]
,
\end{align}
with the temperature $T$ and a normalized sum over momenta $\bvec{k}$ in the first Brillouin zone. Furthermore,
$F_\imp[\bvec{\Delta}]$ is the free-energy functional of a multiorbital Anderson impurity model with a hybridization function $\bvec{\Delta}$, having the same local interaction as the multiorbital Hubbard model at hand. 

We can eliminate arguments from $\Gamma$ by using stationary conditions.
Following Ref.~\cite{Kotliar1999}, we want to analyze the Mott transition via the frequency-dependent order parameter $\bvec{\Delta}$.
Hence, we in turn eliminate $\bvec{G}_\loc$ and $\bvec{S}$ from $\Gamma$. First, we use
\begin{align}
\frac{\delta \Gamma}{\delta \bvec{G}_\loc} \overset{!}{=} 0
\quad \Rightarrow \quad
\bvec{G}_\loc^{-1} = \bvec{S} - \bvec{\Delta}
,
\label{eq:dGamma_dG}
\end{align}
so that $\Gamma$ becomes
\begin{align}
\Gamma[\bvec{S}, \bvec{\Delta}] 
= 
F_\imp[\bvec{\Delta}]
+ \mathrm{Tr}_{\mi\nu} \ln [ \bvec{S} - \bvec{\Delta} ] 
- \mathrm{Tr}_{\mi\nu,\bvec{k}} \ln [ \bvec{S} - \tilde{\bvec{h}}_\bvec{k} ]
.
\label{eq:Gamma_DS}
\end{align}
Second, we have
\begin{subequations}
\begin{align}
\frac{\delta \Gamma}{\delta \bvec{S}} \overset{!}{=} 0 
\quad \Rightarrow \quad
& (\bvec{S} - \bvec{\Delta})^{-1} 
= 
\sum_\bvec{k} (\bvec{S} - \tilde{\bvec{h}}_\bvec{k})^{-1}
\label{eq:dGamma_dS}
,
\\
& \bvec{\Delta} 
= 
\bvec{S} - \Big[ \sum_\bvec{k} (\bvec{S} - \tilde{\bvec{h}}_\bvec{k})^{-1} \Big]^{-1}
.
\label{eq:dGamma_dS_for_Delta}
\end{align}
\end{subequations}
The last relation defines $\bvec{S}[\bvec{\Delta}]$ by inversion.
Thereby, we get
\begin{align}
\Gamma[\bvec{\Delta}] 
& = 
F_\imp[\bvec{\Delta}]
+ \mathrm{Tr}_{\mi\nu} \ln [ \bvec{S}[\bvec{\Delta}] - \bvec{\Delta} ] 
\nonumber
\\
& \
- \mathrm{Tr}_{\mi\nu,\bvec{k}} \ln [ \bvec{S}[\bvec{\Delta}] - \tilde{\bvec{h}}_\bvec{k} ]
.
\label{eq:Gamma_D}
\end{align}
This is our desired expression for $\Gamma$ as a functional of $\bvec{\Delta}$. 

Before analyzing $\Gamma[\bvec{\Delta}]$ around the Mott transition, we briefly check that the third and last stationary condition reproduces the DMFT self-consistency condition.
Using $\delta F_\imp[\bvec{\Delta}] / \delta \bvec{\Delta} = T \bvec{g}$,
where $\bvec{g}$ is the full propagator of the impurity, yields
\begin{align}
\frac{\delta \Gamma}{\delta \bvec{\Delta}} \overset{!}{=} 0 
\quad \Rightarrow \quad
\bvec{g} = ( \bvec{S} - \bvec{\Delta} )^{-1}
,
\label{eq:dGamma_dD}
\end{align}
since the terms with $\bvec{S}'[\bvec{\Delta}]$ cancel by virtue of Eq.~\eqref{eq:dGamma_dS}.
Combining Eqs.~\eqref{eq:dGamma_dG}, \eqref{eq:dGamma_dS}, and \eqref{eq:dGamma_dD}, we indeed obtain
\begin{align}
(\bvec{S} - \bvec{\Delta})^{-1}
=
\bvec{g} 
\overset{!}{=} 
\bvec{G}_\loc
= 
\sum_\bvec{k} (\bvec{S} - \tilde{\bvec{h}}_\bvec{k})^{-1}
,
\end{align}
i.e., the DMFT self-consistency condition between impurity and lattice model.

Now, we go back to Eq.~\eqref{eq:Gamma_D}.
We want to expand $\Gamma[\bvec{\Delta}]$ for a special point in the phase diagram, 
namely close to the transition from the Mott phase, where all orbitals are insulating,
to a (at least partially) metallic phase.
We will focus on low frequencies.
In the Mott phase, each component of $\bvec{\Delta}$ vanishes at low frequencies 
whereas the components of $\bvec{S}$, containing the self-energy, diverge.
Indeed, the relation between $\bvec{S}$ and $\bvec{\Delta}$ is given
by Eq.~\eqref{eq:dGamma_dS}, which can be written as
\begin{align}
\big( 1 - \bvec{\Delta} \bvec{S}^{-1} \big)^{-1}
& =
\sum_\bvec{k} \big( 1 - \tilde{\bvec{h}}_\bvec{k} \bvec{S}^{-1} \big)^{-1}
.
\end{align}
Expanding each side to the first nonvanishing order yields
\begin{align}
\bvec{\Delta} 
& =
\sum_\bvec{k}
\tilde{\bvec{h}}_\bvec{k} \bvec{S}^{-1} \tilde{\bvec{h}}_\bvec{k} 
+ \mathit{O} \big( \bvec{S}^{-2} \big)
,
\label{eq:Delta_S_exp}
\end{align}
as $\sum_\bvec{k} \tilde{\bvec{h}}_\bvec{k} \!=\! 0$.
We see that $\bvec{\Delta}$ and $\bvec{S}^{-1}$ are reciprocally related
in our expansion,
with factors of $\tilde{\bvec{h}}_\bvec{k}$ ensuring the correct dimensions.
The expansion of $\Gamma[\bvec{\Delta}]$ from Eq.~\eqref{eq:Gamma_D} follows as
\begin{align}
& 
\Gamma[\bvec{\Delta}] 
- 
F_\imp[\bvec{\Delta}]
\nonumber
\\
\nonumber
& \ = 
\mathrm{Tr}_{\mi\nu} \ln \big[ 1 - \bvec{\Delta} \bvec{S}^{-1} \big]
- \mathrm{Tr}_{\mi\nu,\bvec{k}} \ln \big[ 1 - \tilde{\bvec{h}}_\bvec{k} \bvec{S}^{-1} \big]
\\
\nonumber
& \ =
- \mathrm{Tr}_{\mi\nu} \big[ \bvec{\Delta} \bvec{S}^{-1} \big]
+ \tfrac{1}{2} \mathrm{Tr}_{\mi\nu,\bvec{k}} \big[ \tilde{\bvec{h}}_\bvec{k} \bvec{S}^{-1} \tilde{\bvec{h}}_\bvec{k} \bvec{S}^{-1} \big]
+ \mathit{O} \big( \bvec{S}^{-3} \big)
\\
& \ =
- \tfrac{1}{2} \mathrm{Tr}_{\mi\nu} \big[ \bvec{\Delta} \bvec{S}^{-1} \big]
+ \mathit{O} \big( \bvec{\Delta}^3 \big)
,
\end{align}
where we employed Eq.~\eqref{eq:Delta_S_exp} in the last step.

To find an explicit expression of $\bvec{S}[\bvec{\Delta}]$ for small $\bvec{\Delta}$,
we can rewrite Eq.~\eqref{eq:Delta_S_exp} as a matrix-vector product:
\begin{align}
\Delta^n
\approx
\sum_\bvec{k} \sum_m \tilde{h}_\bvec{k}^{nm} \frac{1}{S^m} \tilde{h}_\bvec{k}^{mn}
\quad \Leftrightarrow \quad
\vec{\Delta} \approx \overline{\bvec{t}^2} \vec{S}_\mathrm{inv}
,
\label{eq:Delta_S_vec}
\end{align}
where the diagonal matrices $\bvec{\Delta}$ and $\bvec{S}^{-1}$ were cast into vectors,
with $(\vec{\Delta})^n = \Delta^n$ and 
$(\vec{S}_\mathrm{inv})^n = 1/S^n$,
and we defined
\begin{align}
\big[\, \overline{\bvec{t}^2} \, \big]^{nm}
& =
\sum_\bvec{k} \big| \tilde{h}_\bvec{k}^{nm} \big|^2
.
\label{eq:t2bar}
\end{align}
This allows us to invert Eq.~\eqref{eq:Delta_S_vec} as 
$\vec{S}_\mathrm{inv} \approx \big( \overline{\bvec{t}^2} \big)^{-1} \vec{\Delta}$,
so that
\begin{align}
\mathrm{Tr} \big[ \bvec{\Delta} \bvec{S}^{-1} \big]
=
\vec{\Delta}^T \vec{S}_\mathrm{inv}
=
\vec{\Delta}^T \big( \overline{\bvec{t}^2} \big)^{-1} \vec{\Delta}
+
\mathit{O} (\bvec{\Delta}^3)
,
\end{align}
where $\vec{\Delta}^T$ is the transpose of $\vec{\Delta}$.
We thus obtain our final result
\begin{align}
\Gamma[\bvec{\Delta}] 
=
F_\imp[\bvec{\Delta}]
- 
\tfrac{1}{2} T \sum_{\mi\nu} \vec{\Delta}^T \big(\overline{\bvec{t}^2} \big)^{-1} \vec{\Delta}
+
\mathit{O} (\bvec{\Delta}^3)
.
\label{eq:GammaD_final}
\end{align}

Let us briefly check the stationary condition $\delta \Gamma / \delta \Delta \overset{!}{=} 0$
from the expanded $\Gamma[\bvec{\Delta}]$.
As $\overline{\bvec{t}^2}$ from Eq.~\eqref{eq:t2bar} is symmetric, we get
\begin{align}
\vec{g} = \big(\overline{\bvec{t}^2} \big)^{-1} \vec{\Delta}
\quad \Leftrightarrow \quad
\vec{\Delta} = \overline{\bvec{t}^2} \vec{g}
,
\label{eq:stationary_exp}
\end{align}
with $(\vec{g})^n = g^n$.
In more explicit notation, this means that
\begin{align}
\Delta^n_{\mi\nu}
& =
\sum_\bvec{k} \big| \tilde{h}_\bvec{k}^{nm} \big|^2 g^m_{\mi\nu}
=
\sum_\bvec{k} \tilde{h}_\bvec{k}^{nm} \frac{1}{S^m_{\mi\nu}} \tilde{h}_\bvec{k}^{mn}
+ 
\mathit{O}\big( \bvec{S}_{\mi\nu}^{-2} \big)
\nonumber
\\
& =
\bigg[
\sum_\bvec{k} \tilde{\bvec{h}}_\bvec{k} \bvec{G}_{\bvec{k},\mi\nu} \tilde{\bvec{h}}_\bvec{k} 
\bigg]_{nn}
+ 
\mathit{O}\big( \bvec{S}_{\mi\nu}^{-2} \big)
.
\end{align}
Hence, we recover the result from Eq.~\eqref{eq:Delta_G_matrix} if taken close to the Mott phase,
where all self-energy components diverge at low frequencies.

The interested reader may note that Eq.~\eqref{eq:GammaD_final}
resembles the single-orbital expression 
$\Gamma[\Delta] 
= F_\imp[\Delta] - \tfrac{1}{2} T \sum_{\mi\nu} \Delta^2/t^2$,
which is the DMFT free-energy functional---without further 
approximation---on the Bethe lattice \cite{Kotliar1999}.
(Indeed, $\delta \Gamma / \delta \Delta \overset{!}{=} 0$ reproduces the well-known self-consistency condition 
$\Delta = t^2 g$ on the Bethe lattice \cite{Georges1996}.)
By contrast, Eq.~\eqref{eq:GammaD_final} holds for multiple orbitals on any lattice, 
but only to second order in $\bvec{\Delta}$.

Equation~\eqref{eq:GammaD_final} gives the DMFT free-energy functional of a general multiorbital Hubbard model close to the Mott transition.
Just as for the single-orbital case \cite{Kotliar1999},
the low-frequency hybridization function can be used as an order parameter for the Mott transition.
Insulating behavior corresponds to 
$\Ac_{\Delta,\nu=0} = 0$ (or $\Im \Delta_{\mi\nu \to 0} = 0$)
and metallic behavior to 
$\Ac_{\Delta,\nu=0} \neq 0$ (or $\Im \Delta_{\mi\nu \to 0} \neq 0$)
for sufficiently low $T$.
This can be generalized to the multiorbital case:
in the Mott phase, $\Ac^n_{\Delta,\nu=0} = 0$ for all orbitals $n$;
in the OSMP, $\Ac^n_{\Delta,\nu=0} = 0$ and $\Ac^m_{\nu=0} \neq 0$
for some $n$ and $m \neq n$; 
for a Fermi liquid, $\Ac^n_{\Delta,\nu=0} \neq 0$ for all $n$.

In the presence of interorbital hopping, $\overline{\bvec{t}^2}$ 
from Eq.~\eqref{eq:t2bar}
and its inverse have off-diagonal elements. 
Consequently, 
the last term of Eq.~\eqref{eq:GammaD_final} yields an overall quadratic term in the free energy 
that couples the hybridization function of different orbitals. 
Due to this coupling, the order parameters of the different orbitals are no longer independent,
but one ``drags'' the other along.
If, at sufficiently low $T$,
$\Delta^n_{\mi\nu}$ for some $n$ exhibits a metallic low-frequency behavior,
another $\Delta^m_{\mi\nu}$ ($m \neq n$) connected by interorbital hopping will follow
and show similar metallic behavior.

Going back to our two-orbital system, 
we can make this precise.
We label the diagonal elements of $\overline{\bvec{t}^2}$ from Eq.~\eqref{eq:t2bar}
with heavy ($\hv$) and light ($\lt$) and the off-diagonal elements
as interorbital ($\io$). 
Note that all elements of $\overline{\bvec{t}^2}$ are positive.
Inversion of the $2 \!\times\! 2$ matrix 
in Eq.~\eqref{eq:GammaD_final} yields
\begin{align}
& \Gamma[\bvec{\Delta}] 
\approx
F_\imp[\bvec{\Delta}]
- 
\frac{1}{2}
\frac{1}{ \big(\overline{t^2}\big)_{\mathrm{lt}} \big(\overline{t^2}\big)_{\mathrm{hv}} 
- \big(\overline{t^2}\big)_{\mathrm{io}} \big(\overline{t^2}\big)_{\mathrm{io}} }
%\nonumber
%\\
%& \ \times
%T
%\sum_{\mi\nu, n=\mathrm{lt},\mathrm{hv}}
%\!\! \Big[ \Delta^n_{\mi\nu} \Delta^n_{\mi\nu} \big(\overline{t^2}\big)_m - 
%\Delta^n_{\mi\nu} \Delta^m_{\mi\nu} \big(\overline{t^2}\big)_{\mathrm{io}} \Big]
%\bigg|_{n \neq m}
\nonumber
\\
& \ \times
T
\sum_{\mi\nu}
\!\! \Big[ 
(\Delta^\lt_{\mi\nu})^2 \big(\overline{t^2}\big)_\hv
+
(\Delta^\hv_{\mi\nu})^2 \big(\overline{t^2}\big)_\lt
- 
2\Delta^\lt_{\mi\nu} \Delta^\hv_{\mi\nu} \big(\overline{t^2}\big)_{\mathrm{io}} \Big]
.
\label{eq:Gamma_D_2orb}
\end{align}
The first two terms in the bracket give the \textit{cost} of forming 
$\Delta^n_{\mi\nu}$ (the ``Weiss field'') \cite{Kotliar1999},
i.e., forming a metallic hybridization with
$\Im \Delta^n_{\mi\nu \to 0} \neq 0$.
Indeed, at half filling, $\Delta^n_{\mi\nu}$ is purely imaginary and these terms come with a positive sign.
By contrast, the last term gives an energy \textit{gain}
(negative sign for purely imaginary $\Delta^n_{\mi\nu}$)
by having similar hybridization functions in both orbitals,
thus maximizing 
$T \sum_{\mi\nu} \Delta^\lt_{\mi\nu} \Delta^\hv_{\mi\nu}$.
Indeed, if we leave the Mott phase through metallic behavior in the light orbital
($\Im \Delta^\lt_{\mi\nu \to 0} \neq 0$),
then the last term of Eq.~\eqref{eq:Gamma_D_2orb} 
prefers a similarly metallic heavy orbital, since 
$\Im \Delta^\lt_{\mi\nu \to 0} \Im \Delta^\hv_{\mi\nu \to 0} \neq 0$
gives an energy gain compared to
$\Im \Delta^\lt_{\mi\nu \to 0} \Im \Delta^\hv_{\mi\nu \to 0} = 0$.
While $F_\imp[\bvec{\Delta}]$ in the multiorbital case is difficult to analyze explicitly,
the numerical solution of the DMFT equations as in the main text confirms this picture.

\end{document}